\documentclass[aps,prxquantum,twocolumn,superscriptaddress,nofootinbib]{revtex4}

\usepackage{graphicx}
\usepackage{dcolumn}
\usepackage{bm}
\usepackage{amssymb}
\usepackage{amsfonts}
\usepackage{latexsym}
\usepackage{enumerate}
\usepackage[dvipdfmx]{color} 
\usepackage{amsmath}
\usepackage[dvipdfmx]{epsfig}
\usepackage{times}
\usepackage{algorithm}
\usepackage{algorithmic}
\usepackage{cases}

\newtheorem{definition}{Definition}

\begin{document}
\widetext
\title{Unconditional verification of quantum computation with classical light}
\author{Yuki Takeuchi}
\email{yuki.takeuchi@ntt.com}
\affiliation{NTT Communication Science Laboratories, NTT Corporation, 3-1 Morinosato Wakamiya, Atsugi, Kanagawa 243-0198, Japan}
\affiliation{NTT Research Center for Theoretical Quantum Information, NTT Corporation, 3-1 Morinosato Wakamiya, Atsugi, Kanagawa 243-0198, Japan}
\author{Akihiro Mizutani}
\email{mizutani@eng.u-toyama.ac.jp}
\affiliation{Faculty of Engineering, University of Toyama, Gofuku 3190, Toyama 930-8555, Japan}

\begin{abstract}
Verification of quantum computation is a task to efficiently check whether an output given from a quantum computer is correct.
Existing verification protocols conducted between a quantum computer to be verified and a verifier necessitate quantum communication to unconditionally detect any malicious behavior of the quantum computer solving any promise problem in ${\sf BQP}$.
In this paper, we remove the necessity of the communication of qubits by proposing a ``physically classical" verification protocol in which the verifier just sends coherent light to the quantum computer.
\end{abstract}
\maketitle

\section{Introduction}
Quantum computers are expected to outperform classical computers in a variety of applications, from cryptanalysis~\cite{BHT97,S97,HSTX20} to physics~\cite{K95,NPdB21,CDBBT24} and chemistry~\cite{KJLMA08,GL22,CFGHLMW23}.
The flip side of their advantages is their susceptibility to noises. 
Therefore, to obtain benefits from quantum computers, it is necessary to devise an efficient protocol for checking whether a quantum computer outputs a correct answer, a task called verification of quantum computation~\cite{GKK19,EHWRMPCK20,KR21}.
One may think that verification protocols would become useless if a sufficient number of qubits for quantum error correction~\cite{G20} can be created; however, this is not the case, because it will still be necessary to check whether an implemented quantum error correction scheme faithfully works.
Multiple small-scale experiments~\cite{BFKW13,ZZCPXYYYHXCLG20,JWQCCLXSZM20,FMMD21} have already demonstrated progress toward the realization of verifiable quantum information processing.

Verification protocols are evaluated in terms of five properties: (i) whether the soundness is information theoretic or computational; (ii) the type of communication required for a verifier; (iii) the number of necessary non-communicating provers, which are quantum computers to be verified; (iv) the presence or absence of a trusted third party who assuredly follows procedures in the protocols; and (v) the problems to which the protocols can be applied.
Here, soundness can be information theoretic or computational. Information-theoretic soundness means that any malicious computationally unbounded prover who outputs incorrect answers can be rejected.
On the other hand, computational soundness is a property for rejecting malicious quantum polynomial-time provers.
Hence, the former property is stronger than the latter.
As for (iii), it would, in general, be hard to guarantee that multiple provers do not communicate with each other.
Therefore, the ultimate goal is to devise a protocol such that (i) the soundness is information theoretic, i.e., even if the quantum computer outputs an incorrect answer by taking superpolynomial time, it can be properly detected, (ii) the classical communication is sufficient, (iii) \& (iv) a single prover is sufficient, and (v) the protocol is applicable to any problem in ${\sf BQP}$, which is a set of promise problems (i.e., problems that can be answered by YES or NO) solvable in quantum polynomial time.
However, it is hard to construct such an outstanding protocol with currently known theoretical techniques.
Although its impossibility has not been shown, which immediately implies ${\sf P}\neq{\sf PSPACE}$, some existing results~\cite{FHM18,MT20} have revealed the difficulty of its construction.

\begin{figure*}[t]
\includegraphics[width=17cm, clip]{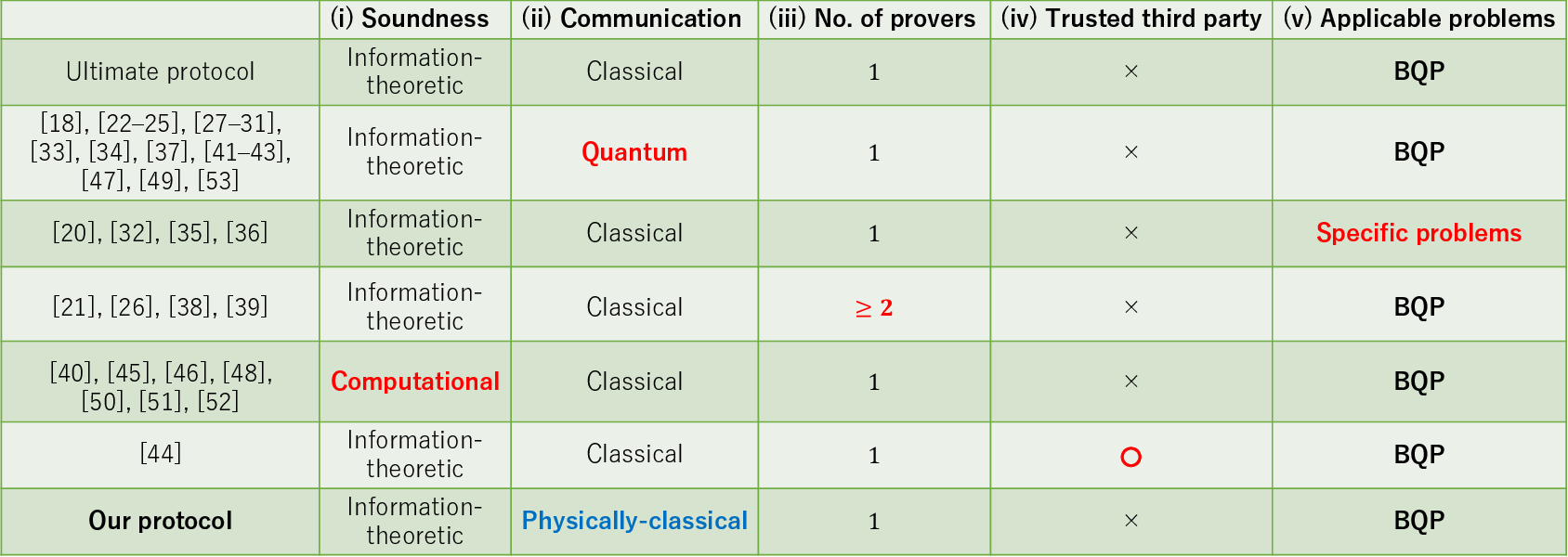}
\caption{Summary of existing verification protocols. The third column indicates the type of communication required for the verifier. In the fifth column, the cross and circle marks mean that a trusted third party is unnecessary and necessary, respectively. In the sixth column, {\bf BQP} means that the protocols can be applied to any problem in ${\sf BQP}$. The disadvantages of the verification protocols are highlighted in bold red. Our contribution is emphasized in bold blue.}
\label{summary}
\end{figure*}

This situation has led to the development of numerous verification protocols~\cite{M12,RUV13,M14,GKW15,HM15,MNS16,M16,HKSE17,KW17,FK17,FHM18,M18,DOF18,TM18,B18,FKD18,LMNT18,MTN18,TMMMF19,CGJV19,G19,GV19,FKD19,LDTAF19,ZH19,M20,ACGH20,CCY20,LMKO21,CLLW22,KKLMO22,M22,Z22,GMP23,LZH23}, each of which has its own advantages and disadvantages.
These protocols fall into five types of approaches~\cite{FN1} as summarized in Fig.~\ref{summary}.
In terms of practicality, we stick to achieving the information-theoretic soundness and keeping the total number of provers and a trusted third party one.
This is because it would be unclear how to guarantee that the prover's computation is completed in polynomial time, that the multiple provers do not communicate with each other, and that a third party does not cooperate with a prover.
In this sense, from Fig.~\ref{summary}, quantum communication is necessary to verify any problem in ${\sf BQP}$ with existing practical verification protocols.

In this paper, to remove the necessity of qubit communication, we propose a ``physically classical" verification protocol in which the verifier just sends coherent light to the quantum computer to be verified.
The transmission of coherent light is the same as classical communication in the sense that classical bits are sent by using light in the real world, while they are absolutely different from the viewpoint of information theory.
To obtain our protocol, we first modify the verification protocol in Ref.~\cite{M20} such that a trusted third party is combined with the verifier, and hence qubit communication becomes necessary for the verifier.
Then, to remove the qubit communication, we combine it with the technique of using coherent light as a substitute for qubits.
This technique has been developed in several quantum information processing tasks such as quantum key distribution (QKD)~\cite{B92,HIGM95,IWY03,LMC05,LP07}, blind quantum computation~\cite{DKL12}, and the demonstration of quantum advantage~\cite{CKDK21}.
All the approaches except for that in Ref.~\cite{DKL12} were developed for the transmission of classical bits.
For example, random single-qubit states chosen from $\{|0\rangle,|1\rangle,|\pm\rangle\}$ with $|\pm\rangle\equiv(|0\rangle+|1\rangle)/\sqrt{2}$ are transmitted to share a secret bit string between two parties in the original QKD protocol~\cite{BB84}.
This quantum communication is replaced with coherent light communication in Refs.~\cite{B92,HIGM95,IWY03,LMC05,LP07}.
However, the transmission of classical bits is insufficient for existing verification protocols, which is why we use the technique in Ref.~\cite{DKL12}.
Moreover, since its proof of principle experiment has already been demonstrated over a distance of 100 km fiber~\cite{JWHXSZZLYWLXZP19}, its use is preferable from a practical point of view.

The remaining issue to be resolved is that the technique in Ref.~\cite{DKL12} seems to be incompatible with existing verification protocols.
The technique replaces the transmission of single-qubit states in a single plane of the Bloch sphere with that of coherent light, and it works when the single-qubit states are chosen uniformly at random.
Although the single-qubit states used in the protocol of Ref.~\cite{M20} are in the $x$-$z$ plane of the Bloch sphere, these qubits have the same random basis, i.e., the basis is not random among them.
To fill in this gap, we further modify the protocol in Ref.~\cite{M20} (for details, see Protocol 1).

\section{Verification protocol with quantum communication}
\label{II}
As the first step to constructing our physically classical verification protocol, we propose a verification protocol with communication of qubits by modifying the protocol in Ref.~\cite{M20}.
The purpose of our protocol is to verify quantum computation solving any problem in ${\sf BQP}$:
\begin{definition}[\cite{BV97}]
A promise problem $L=(L_{\rm yes},L_{\rm no})$ is in ${\sf BQP}$ if and only if there exists a uniform family $\{U_x\}_x$ of polynomial-size quantum circuits such that when $x\in L_{\rm yes}$, $\langle 0^n|U_x^\dag(|1\rangle\langle 1|\otimes I^{\otimes n-1})U_x|0^n\rangle\ge 2/3$ and when $x\in L_{\rm no}$, $\langle 0^n|U_x^\dag(|1\rangle\langle 1|\otimes I^{\otimes n-1})U_x|0^n\rangle\le 1/3$.
Here, $I\equiv|0\rangle\langle 0|+|1\rangle\langle 1|$ is the two-dimensional identity operator, $n$ is a polynomial in $|x|$, and $|x|$ is the length of the instance $x$.
\end{definition}
Simply speaking, ${\sf BQP}$ is a set of problems that can be efficiently solved with a universal quantum computer.

The protocol in Ref.~\cite{M20} is based on the local Hamiltonian problem~\cite{KSV02}.
Let $L$ be a promise problem in ${\sf BQP}$.
For any instance $x\in L$, we define the $N$-qubit Hamiltonian~\cite{FN2}
\begin{eqnarray}
\nonumber
H_x&\equiv&\sum_{1\le i<j\le N}\cfrac{p_{ij}^{(x)}}{2}\times\\
&&\left(\cfrac{I^{\otimes N}+c_{ij}^{(x)}X_i\otimes X_j}{2}+\cfrac{I^{\otimes N}+c_{ij}^{(x)}Y_i\otimes Y_j}{2}\right),\ \ \ \ \
\end{eqnarray}
where for all $i$, $j$, and $x$, $p_{ij}^{(x)}\ge0$, $\sum_{i<j}p_{ij}^{(x)}=1$, $c_{ij}^{(x)}\in\{1,-1\}$, and $X_i$ and $Y_i$ represent the Pauli-$X$ and $Y$ operators applied to the $i$th qubit, respectively.
Note that $N$ is a polynomial in $|x|$.
Let us assume that the prover declares $x\in L_{\rm yes}$, i.e., the quantum computer outputs YES as the correct answer.
We also define certain non-negative values $a$ and $b$ such that $1\ge b-a\ge 1/f(|x|)$ for a polynomial function $f$.
From the ${\sf BQP}$-hardness (more precisely, ${\sf QMA}$-completeness) of the $2$-local Hamiltonian problem~\cite{KKR06,BL08,CM16}, the verifier can efficiently find $\{p_{ij}^{(x)},c_{ij}^{(x)}\}_{1\le i<j\le N}$ such that (i) when the prover is honest (i.e., the correct answer is indeed YES), there exists an efficiently preparable quantum state $|\eta\rangle$ whose energy $\langle\eta|H_x|\eta\rangle$ is at most $a$, and (ii) when the prover is malicious (i.e., the correct answer is NO), the ground-state energy is at least $b$.
Since ${\sf BQP}$ is closed under complement, even when the prover declares $x\in L_{\rm no}$, the verifier can efficiently find $\{p_{ij}^{(x)},c_{ij}^{(x)}\}_{1\le i<j\le N}$ having the same property.
The verifier in the protocol of Ref.~\cite{M20} decides whether the prover is honest or malicious by measuring the energy of $H_x$ with the aid of a trusted third party.

With the above idea in mind, we modify the protocol in Ref.~\cite{M20} as follows:

\medskip
\noindent{\bf [Protocol 1]}
\begin{enumerate}
\item The verifier chooses two tuples $(h_1,\ldots,h_N)\in\{0,1\}^N$ and $(s_1,\ldots,s_N)\in\{0,1\}^N$ uniformly at random.
Then the verifier sends
\begin{eqnarray}
|\psi_V\rangle\equiv\bigotimes_{i=1}^N\left(S^{h_i}H|s_i\rangle\right)
\end{eqnarray}
to the prover, where $S\equiv|0\rangle\langle 0|+i|1\rangle\langle 1|$ is the $S$ gate, and $H\equiv|+\rangle\langle 0|+|-\rangle\langle 1|$ is the Hadamard gate.
\item The prover performs a POVM measurement $\{\Pi_{wz}\}_{w,z\in\{0,1\}^N}$ on the received state $|\psi_V\rangle$ and sends the measurement outcomes $w$ and $z$ to the verifier.
If the prover is honest, $\{\Pi_{wz}\}_{w,z}$ corresponds to the $N$ Bell measurements on each qubits of $|\psi_V\rangle$ and $|\eta\rangle$.
Therefore, the prover's operation is essentially equivalent to the quantum teleportation of $|\eta\rangle$.
On the other hand, if the prover is malicious, $\{\Pi_{wz}\}_{w,z}$ can be an arbitrary measurement.
\item The verifier samples a set $(i,j)$ with probability $p_{ij}^{(x)}$. Since the cardinality of the set $\{p_{ij}^{(x)}\}_{i<j}$ is $N(N-1)/2$, this sampling can be performed in classical polynomial time in $N$.
If $h_i=h_j$, then the verifier proceeds to the next step.
Otherwise, the verifier accepts the prover.
\item Let $s'_k\equiv s_k\oplus z_k\oplus h_kw_k$ for all $1\le k\le N$, where $z_k$ and $w_k$ are the $k$th bits of $z$ and $w$, respectively.
If $(-1)^{s'_i+s'_j}=-c_{ij}^{(x)}$, the verifier accepts the prover.
Otherwise, the verifier rejects the prover.
\end{enumerate}

There exist two differences between the original protocol in Ref.~\cite{M20} and Protocol 1.
First, a trusted third party is merged with the verifier in step 1.
Second, the bases $\{h_i\}_{i=1}^N$ are randomly chosen for each qubit in Protocol 1, while the basis of all qubits is determined by a single random bit $h\in\{0,1\}$ in the original protocol.
These differences are essential for devising our physically classical verification protocol, which is described in the next section.

As shown in Appendix A, the acceptance probability $p_{\rm acc}$ of Protocol 1 is at least $1-a/2$ (at most $1-b/2$) when the prover is honest (malicious).
Since the gap of $p_{\rm acc}$ in the two cases is
\begin{eqnarray}
\left(1-\cfrac{a}{2}\right)-\left(1-\cfrac{b}{2}\right)=\cfrac{b-a}{2}\ge\cfrac{1}{2f(|x|)},
\end{eqnarray}
the verifier can distinguish between the honest and malicious provers by repeating Protocol 1 in parallel a polynomial number of times.

\section{Verification of quantum computation with coherent light}
The purpose of this section is to replace the quantum communication in Protocol 1 with the transmission of coherent light.
To this end, we use the remote blind qubit state preparation (RBSP) protocol in Ref.~\cite{DKL12}.
The adoption of the RBSP protocol is possible due to the modification in the previous section.

\begin{figure*}[t]
\includegraphics[width=18cm, clip]{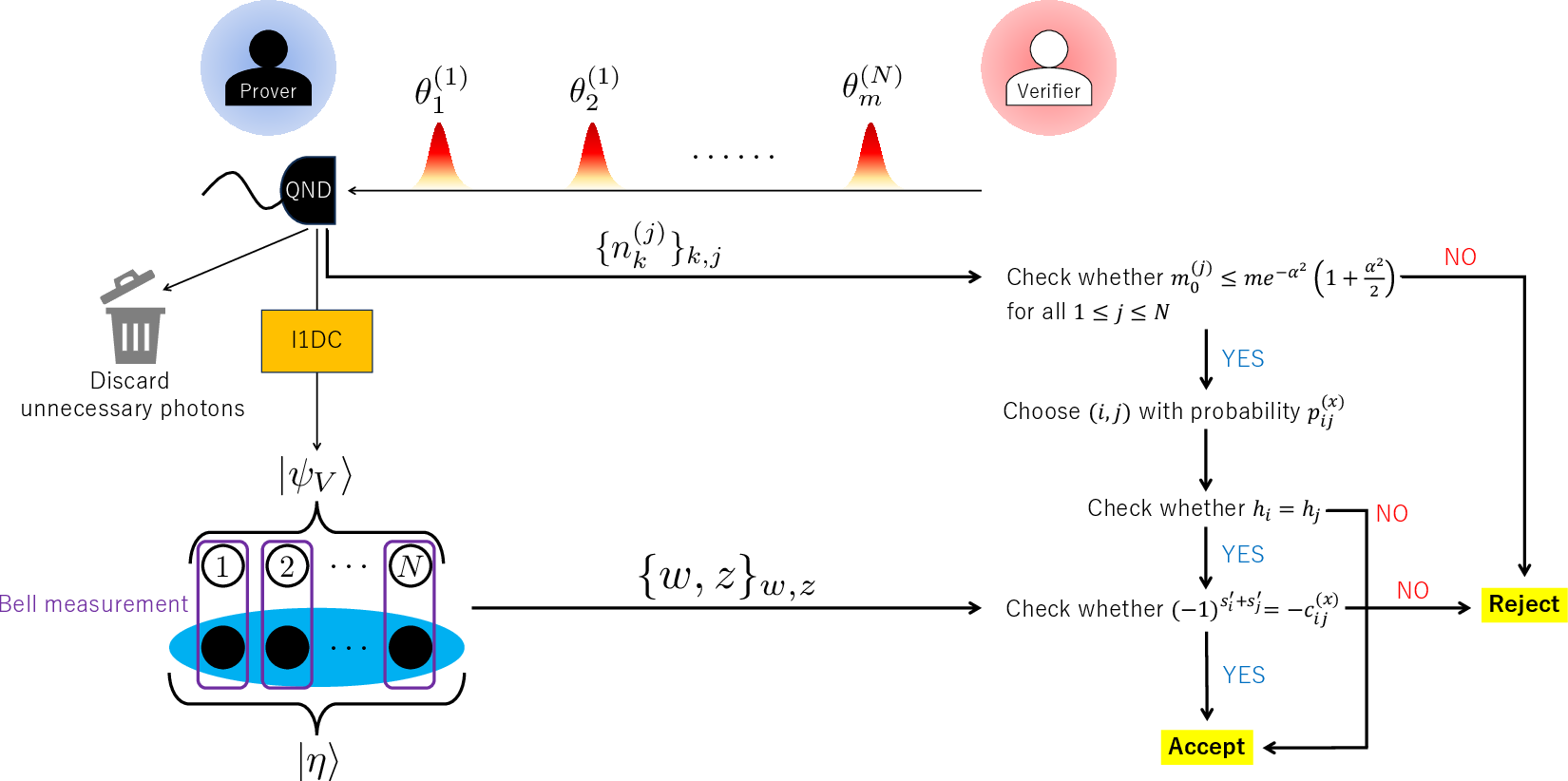}
\caption{Schematic of our verification protocol with an honest prover. Classical communication and operations are represented by bold arrows.
Non-classical communication and operations are represented by thin arrows.
Each purple rectangular enclosure represents the Bell measurement.
The verifier first sends phase-randomized coherent states with randomized polarization angles $\{\theta_k^{(j)}\}_{1\le k\le m,1\le j\le N}\in\{0,\pi/2,\pi,3\pi/2\}^{mN}$ to the prover.
Then the prover performs QND measurements on all the received coherent states and sends the measurement outcomes $\{n_k^{(j)}\}_{k,j}$ to the verifier.
The prover also discards unnecessary photons and performs the I1DC protocol by using the remaining photons (see steps (b) and (e) in Protocol 2).
As a result, the prover obtains an $N$-qubit state $\otimes_{j=1}^N|\psi_V^{(j)}\rangle$, which will be used as $|\psi_V\rangle$ in step 1 of Protocol 1.
The prover performs the quantum teleportation of the low-energy state $|\eta\rangle$ by measuring $N$ pairs of qubits of $|\psi_V\rangle$ and $|\eta\rangle$ in the Bell bases and sending $\{w,z\}_{w,z\in\{0,1\}^N}$ to the verifier.
On the other hand, the verifier checks whether Eq.~(\ref{threshold}) holds for all $1\le j\le N$.
If it does not hold, the verifier rejects the prover.
If it holds, the prover chooses the pair $(i,j)$ with probability $p_{ij}^{(x)}$ and then checks whether $h_i=h_j$ (see step 3 in Protocol 1).
If it is not satisfied, the prover is automatically accepted.
Otherwise, the verifier calculates $(-1)^{s'_i+s'_j}$ by using $w$ and $z$ sent from the prover (see step 4 in Protocol 1).
When it is equal to $-c_{ij}^{(x)}$, the verifier accepts the prover.
On the other hand, if it is not the case, the prover is rejected.}
\label{protocol}
\end{figure*}

We replace step 1 in Protocol 1 with  the following protocol:

\medskip
\noindent{\bf [Protocol 2]}
\begin{enumerate}
\item The verifier and prover conduct the following steps $N$ times.
Note that when the prover is malicious, the prover can apply any completely positive trace-preserving (CPTP) map in each step.
\begin{enumerate}
\item Let $m(\ge8)$ be a natural number specified later.
For the $j$th repetition $(1\le j\le N)$, the verifier chooses a tuple $(\theta_1^{(j)},\ldots,\theta_m^{(j)})\in\{0,\pi/2,\pi,3\pi/2\}^m$ uniformly at random.
Then the verifier sends the $m$ phase-randomized coherent states
\begin{eqnarray}
\label{cprcs}
\bigotimes_{k=1}^m\left(e^{-\alpha^2}\sum_{n=0}^\infty\cfrac{\alpha^{2n}}{n!}|n_{\theta_k^{(j)}}\rangle\langle n_{\theta_k^{(j)}}|\right)
\end{eqnarray}
to the prover, where $(8/m)^{1/4}\le\alpha\le1$, and $|n_\theta\rangle\equiv[(|0\rangle+e^{i\theta}|1\rangle)/\sqrt{2}]^{\otimes n}$ is an $n$-photon state with the polarization angle $\theta$.
Here, $|0\rangle$ and $|1\rangle$ are not photon number states, but the computational basis ones.
\item The prover performs a quantum nondemolition (QND) measurement of the photon number on each of the $m$ coherent states and obtains the measurement outcomes $\{n_k^{(j)}\}_{k=1}^m$.
If $n_k^{(j)}\ge1$, the prover keeps $|1_{\theta_k^{(j)}}\rangle$ at hand and discards the other $(n_k^{(j)}-1)$ photons.
Let $m_0^{(j)}$ be the number of $k$'s such that $n_k^{(j)}=0$.
At the end of this step, the honest prover possesses exactly $(m-m_0^{(j)})$ photons~\cite{FNM1}.
\item The prover sends $\{n_k^{(j)}\}_{k=1}^m$ to the verifier.
\item The verifier calculates $m_0^{(j)}$ from $\{n_k^{(j)}\}_{k=1}^m$.
If
\begin{eqnarray}
\label{threshold}
m_0^{(j)}\le me^{-\alpha^2}\left(1+\cfrac{\alpha^2}{2}\right),
\end{eqnarray}
the verifier and prover proceed to the next step.
Otherwise, the verifier rejects the prover.
\item The prover performs the interlaced 1D cluster computation (I1DC) protocol on the $(m-m_0^{(j)})$ photons as follows: for $l'=1$ to $m-m_0^{(j)}-1$,
\begin{enumerate}
\item Apply $CZ(H\otimes I)$ to the $l'$th and $(l'+1)$th photons, where $CZ\equiv I^{\otimes 2}-2|11\rangle\langle11|$ is the controlled-$Z$ gate.
\item Measure the $l'$th photon in the Pauli-$X$ basis, and obtain the measurement outcome $o_{l'}^{(j)}$.
\end{enumerate}
\item The prover sends the measurement outcomes $\{o_{l'}^{(j)}\}_{l'=1}^{m-m_0^{(j)}-1}$ to the verifier.
The prover keeps the unmeasured $(m-m_0^{(j)})$th photon at hand and will use it as the $j$th qubit $|\psi_V^{(j)}\rangle$ in step 2 of Protocol 1.
\end{enumerate}
\end{enumerate}

By replacing step 1 in Protocol 1 with Protocol 2, we can devise our physically classical verification protocol (see Fig.~\ref{protocol}).
To evaluate its performance, we first derive its lower bound on the acceptance probability $p_{\rm acc}$ for the honest prover as a function of $\alpha$ and $m$.
In step (b) of Protocol 2, the honest prover measures
\begin{eqnarray}
\label{coherentstate}
e^{-\alpha^2}\sum_{n=0}^\infty\cfrac{\alpha^{2n}}{n!}|n_{\theta_k^{(j)}}\rangle\langle n_{\theta_k^{(j)}}|
\end{eqnarray}
in the Fock basis $\{|n\rangle\}_{n\in\mathbb{Z}_{\ge 0}}$.
From Eq.~(\ref{coherentstate}), the probability of obtaining $n_k^{(j)}=0$ is $e^{-\alpha^2}$ for any $k$ and $j$, and hence the mean value of $m_0^{(j)}$ is $me^{-\alpha^2}$.
The Hoeffding inequality~\cite{H63} implies that the probability of the honest prover satisfying Eq.~(\ref{threshold}) is
\begin{eqnarray}
\nonumber
&&{\rm Pr}\left[m_0^{(j)}\le me^{-\alpha^2}\left(1+\cfrac{\alpha^2}{2}\right)\right]\\
&\ge&1-{\rm Pr}\left[m_0^{(j)}-me^{-\alpha^2}\ge me^{-\alpha^2}\cfrac{\alpha^2}{2}\right]\\
\label{passh1}
&\ge&1-{\rm exp}\left(-\cfrac{me^{-2\alpha^2}\alpha^4}{2}\right).
\end{eqnarray}
Furthermore, when Eq.~(\ref{threshold}) is satisfied,
 \begin{eqnarray}
 m-m_0^{(j)}&\ge&m-m\left(1-\cfrac{\alpha^2}{2}\right)\left(1+\cfrac{\alpha^2}{2}\right)\\
 &=&m\cfrac{\alpha^4}{4}\ge 2,
\end{eqnarray}
and hence the prover can definitely perform the I1DC protocol in step (e) of Protocol 2.

For any $1\le l\le m-m_0^{(j)}$, let $\sigma_l^{(j)}\in\{0,\pi/2,\pi,3\pi/2\}$ be the polarization angle of the $l$th remaining photon at the end of step (b) in Protocol 2, i.e., the state of the $l$th remaining photon is $(|0\rangle+e^{i\sigma_l^{(j)}}|1\rangle)/\sqrt{2}$~\cite{FNM3}.
As shown in Ref.~\cite{DKL12}, when the input states are $\{(|0\rangle+e^{i\sigma_l^{(j)}}|1\rangle)/\sqrt{2}\}_{l=1}^{m-m_0^{(j)}}$, the I1DC protocol outputs the measurement outcomes $\{o_{l'}^{(j)}\}_{l'=1}^{m-m_0^{(j)}-1}\in\{0,1\}^{m-m_0^{(j)}-1}$ and the single-qubit state $(|0\rangle+e^{i\varphi^{(j)}}|1\rangle)/\sqrt{2}$, where
\begin{eqnarray}
\varphi^{(j)}\equiv\sum_{l=1}^{m-m_0^{(j)}-1}(-1)^{\sum_{l'=l}^{m-m_0^{(j)}-1}o_{l'}^{(j)}}\sigma_l^{(j)}+\sigma_{m-m_0^{(j)}}^{(j)}.
\end{eqnarray}
Therefore, the verifier can calculate the value of $\varphi^{(j)}$ from the measurement outcomes $\{o_{l'}^{(j)}\}_{l'=1}^{m-m_0^{(j)}-1}$ and the polarization angles $\{\sigma_l^{(j)}\}_{l=1}^{m-m_0^{(j)}}$ in classical polynomial time in $m$.
The value of $\varphi^{(j)}$ is chosen from $\{0,\pi/2,\pi,3\pi/2\}$ with the same probability, $1/4$.
This is because the verifier chooses the value of $\sigma_{m-m_0^{(j)}}^{(j)}$ uniformly at random in step (a) of Protocol 2.
From the above argument, the output state of the I1DC protocol can be expressed as
\begin{eqnarray}
|\psi_V^{(j)}\rangle&=&S^{h_j}H|s_j\rangle
\end{eqnarray}
by using two random bits $h_j$ and $s_j$.
To be more specific, $(h_j,s_j)$ is $(0,0)$, $(0,1)$, $(1,0)$, or $(1,1)$ when $\varphi^{(j)}$ is $0$, $\pi$, $\pi/2$, or $3\pi/2$, respectively.

In conclusion, if Eq.~(\ref{threshold}) is satisfied for all $1\le j\le N$, the verifier accepts the prover with the same probability as that of Protocol 1, i.e., with probability of at least $1-a/2$.
Thus, from Eq.~(\ref{passh1}), the lower bound on $p_{\rm acc}$ of our physically classical verification protocol is
\begin{eqnarray}
\nonumber
&&\left[1-{\rm exp}\left(-\cfrac{me^{-2\alpha^2}\alpha^4}{2}\right)\right]^N\left(1-\cfrac{a}{2}\right)\\
\label{pacchp}
&\ge&1-\cfrac{a}{2}-N{\rm exp}\left(-\cfrac{me^{-2\alpha^2}\alpha^4}{2}\right)
\end{eqnarray}
when the prover is honest.

We next show the information-theoretic soundness of our physically classical verification protocol.
In other words, we give an upper bound on $p_{\rm acc}$ in the case of the malicious prover.
To this end, we observe that the argument in Ref.~\cite{DKL12} can be applied even in our situation.
As an important point, how many photons are transmitted to the prover is randomly decided following the Poisson distribution, and hence the malicious prover cannot decide it even though any quantum operation is allowed for the malicious prover.
We use this randomness to detect the malicious prover's deviation.

Let $m_0^{(j)}$ and $m_1^{(j)}$ be the actual numbers of $k$'s such that $n_k^{(j)}=0$ and $1$, respectively.
Although the malicious prover may not perform QND measurements in step (b) of Protocol 2, the actual photon number $n_k^{(j)}$ is properly defined because the phase-randomized coherent state in Eq.~(\ref{coherentstate}) is diagonalized in the Fock basis.
Despite the malicious prover's ability to perform any quantum operation, there exist only two cases where
\begin{enumerate}
\item[(i)] for at least a single $j$, the actual number $m_0^{(j)}+m_1^{(j)}$ of vacuum and single-photon states is less than or equal to the threshold value $me^{-\alpha^2}(1+\alpha^2/2)$ in step (d) of Protocol 2, and
\item[(ii)] $m_0^{(j)}+m_1^{(j)}$ is more than the threshold value for all $1\le j\le N$.
\end{enumerate}
Let $p_{\rm (i)}$ and $p_{\rm (ii)}$ denote the probabilities of the first and second cases occurring, respectively.
We also define $p_{{\rm acc}|{\rm (i)}}$ and $p_{{\rm acc}|{\rm (ii)}}$ as probabilities of the prover being accepted in the first and second cases, respectively.
By using these notations, the acceptance probability $p_{\rm acc}$ of our physically classical verification protocol is
\begin{eqnarray}
\label{paccnew1}
p_{\rm acc}&=&p_{{\rm acc}|{\rm (i)}}p_{\rm (i)}+p_{{\rm acc}|{\rm (ii)}}p_{\rm (ii)}\le p_{\rm (i)}+p_{{\rm acc}|{\rm (ii)}}.
\end{eqnarray}
We derive an upper bound on $p_{\rm acc}$ by evaluating  $p_{\rm (i)}$ and $p_{{\rm acc}|{\rm (ii)}}$ one by one.

We first evaluate $p_{\rm (i)}$.
From the Hoeffding inequality~\cite{H63}, for all $j$,
\begin{eqnarray}
\nonumber
&&{\rm Pr}\left[m_0^{(j)}+m_1^{(j)}> me^{-\alpha^2}\left(1+\cfrac{\alpha^2}{2}\right)\right]\\
&=&1-{\rm Pr}\left[m_0^{(j)}+m_1^{(j)}\le me^{-\alpha^2}\left(1+\cfrac{\alpha^2}{2}\right)\right]\\
&=&1-{\rm Pr}\left[m\cfrac{1+\alpha^2}{e^{\alpha^2}}-\left(m_0^{(j)}+m_1^{(j)}\right)\ge me^{-\alpha^2}\cfrac{\alpha^2}{2}\right]\ \ \ \ \ \ \ \\
\label{paccm}
&\ge&1-{\rm exp}\left(-\cfrac{me^{-2\alpha^2}\alpha^4}{2}\right),
\end{eqnarray}
where we have used the fact that $m_0^{(j)}+m_1^{(j)}$ converges to $me^{-\alpha^2}(1+\alpha^2)$ to derive the last inequality~\cite{FNM2}.
Therefore,
\begin{eqnarray}
p_{\rm (i)}&\le&1-\left[1-{\rm exp}\left(-\cfrac{me^{-2\alpha^2}\alpha^4}{2}\right)\right]^N\\
\label{paccmp}
&\le&N{\rm exp}\left(-\cfrac{me^{-2\alpha^2}\alpha^4}{2}\right).
\end{eqnarray}

We next evaluate $p_{{\rm acc}|{\rm (ii)}}$.
Let $\tilde{m}_0^{(j)}$ be the number of vacuum states calculated from the measurement outcomes sent by the malicious prover in step (c) of Protocol 2.
The inequality $\tilde{m}_0^{(j)}<m_0^{(j)}+m_1^{(j)}$ has to hold for all $j$ to maximize $p_{{\rm acc}|{\rm (ii)}}$ because the malicious prover is definitely rejected if $\tilde{m}_0^{(j)}\ge m_0^{(j)}+m_1^{(j)}>me^{-\alpha^2}(1+\alpha^2/2)$.
This implies that the input of the I1DC protocol must include at least a single state whose actual photon number is one or zero.

We first consider the case where a single state whose actual photon number is one is included in the input of the I1DC protocol.
Let it be the $l^\ast$th input state whose polarization angle is $\sigma_{l^\ast}^{(j)}\in\{0,\pi/2,\pi,3\pi/2\}$.
In general, the malicious prover applies any CPTP map to the $(m-\tilde{m}_0^{(j)})$ input states, which are used for the I1DC protocol in the case of an honest prover.
We interpret it as that the $l^\ast$th input state is converted to some quantum state by using the other $(m-\tilde{m}_0^{(j)}-1)$ input states as ancillary states.
This interpretation implies that, immediately before step (e) in Protocol 2, the prover's state is
\begin{eqnarray}
\label{statemp1}
\cfrac{1}{4}\sum_{\sigma_{l^\ast}^{(j)}\in\{0,\pi/2,\pi,3\pi/2\}}\mathcal{E}\left(\cfrac{|0\rangle+e^{i\sigma_{l^\ast}^{(j)}}|1\rangle}{\sqrt{2}}\cfrac{\langle0|+e^{-i\sigma_{l^\ast}^{(j)}}\langle1|}{\sqrt{2}}\right)\ \ \
\end{eqnarray}
with some CPTP map $\mathcal{E}$.
Since the verifier chooses each polarization angle independently in step (a) of Protocol 2, and the actual photon number of the $l^\ast$th input state is one, $\mathcal{E}$ does not depend on $\sigma_{l^\ast}^{(j)}$ but can depend on the other input states' polarization angles $\{\sigma_l^{(j)}\}_{1\le l\le m-\tilde{m}_0^{(j)},l\neq l^\ast}$.
To mimic the honest prover, in steps (e) and (f) of Protocol 2, the malicious prover generates and sends the measurement outcomes $\vec{o}^{\ (j)}\equiv\{o_{l'}^{(j)}\}_{l'=1}^{m-\tilde{m}_0^{(j)}-1}\in\{0,1\}^{m-\tilde{m}_0^{(j)}-1}$ by applying some additional CPTP map to the quantum state in Eq.~(\ref{statemp1}).
For simplicity, we define $|+_\theta\rangle\equiv(|0\rangle+e^{i\theta}|1\rangle)/\sqrt{2}$ for any real number $\theta$.
The malicious prover's state is finally
\begin{eqnarray}
\label{statemp2}
\cfrac{1}{4}\sum_{\sigma_{l^\ast}^{(j)}}\sum_{\vec{o}^{\ (j)}}p_{\sigma_{l^\ast}^{(j)}}(\vec{o}^{\ (j)})\mathcal{E}_{\vec{o}^{\ (j)}}\left(|+_{\sigma_{l^\ast}^{(j)}}\rangle\langle+_{\sigma_{l^\ast}^{(j)}}|\right),
\end{eqnarray}
where $p_{\sigma_{l^\ast}^{(j)}}(\vec{o}^{\ (j)})$ is the probability of outputting $\vec{o}^{\ (j)}$, and $\mathcal{E}_{\vec{o}^{\ (j)}}$ is a quantum operation to be applied when the measurement outcomes are $\vec{o}^{\ (j)}$.
The subscript of $p_{\sigma_{l^\ast}^{(j)}}(\vec{o}^{\ (j)})$ just represents that the probability, in general, depends on $|+_{\sigma_{l^\ast}^{(j)}}\rangle$; it does not mean that the prover's CPTP map is constructed by using the value of $\sigma_{l^\ast}^{(j)}$.
In fact, there exists a CPTP map $\mathcal{F}_1^{(j)}$ such that 
\begin{eqnarray}
\nonumber
&&\mathcal{F}_1^{(j)}\left(|+_{\sigma_{l^\ast}^{(j)}}\rangle\langle+_{\sigma_{l^\ast}^{(j)}}|\right)\\
\label{CPTPprop}
&=&\sum_{\vec{o}^{\ (j)}}p_{\sigma_{l^\ast}^{(j)}}(\vec{o}^{\ (j)})\mathcal{E}_{\vec{o}^{\ (j)}}\left(|+_{\sigma_{l^\ast}^{(j)}}\rangle\langle+_{\sigma_{l^\ast}^{(j)}}|\right)
\end{eqnarray}
for any $\sigma_{l^\ast}^{(j)}$.
This is because the verifier keeps the value of $\sigma_{l^\ast}^{(j)}$ private and the actual photon number of the $l^\ast$th input state is one, i.e., the information of $\sigma_{l^\ast}^{(j)}$ is only contained in $|+_{\sigma_{l^\ast}^{(j)}}\rangle$.
For any $\vec{o}^{\ (j)}$ and $\{\sigma_l^{(j)}\}_{l\neq l^\ast}$, there exist $\sigma\in\{0,\pi/2,\pi,3\pi/2\}$ and $c\in\{1,-1\}$ such that
\begin{eqnarray}
\label{tov}
\varphi^{(j)}=\sigma+c\sigma_{l^\ast}^{(j)}.
\end{eqnarray}
From Eq.~(\ref{tov}), we can replace the variable $\sigma_{l^\ast}^{(j)}$ with $\varphi^{(j)}$ in Eq.~(\ref{statemp2}) as follows:
\begin{eqnarray}
&&\cfrac{1}{4}\sum_{\varphi^{(j)}}\sum_{\vec{o}^{\ (j)}}p_{\varphi^{(j)}}(\vec{o}^{\ (j)})\mathcal{E}_{\vec{o}^{\ (j)}}\left(|+_{\varphi^{(j)}}\rangle\langle+_{\varphi^{(j)}}|\right)\\
\label{statemp2.5}
&=&\cfrac{1}{4}\sum_{\varphi^{(j)}}\sum_{\vec{o}^{\ (j)}}p_{\varphi^{(j)}}(\vec{o}^{\ (j)})\mathcal{E}_{\vec{o}^{\ (j)}}\left(|\psi_V^{(j)}\rangle\langle\psi_V^{(j)}|\right),
\end{eqnarray}
where we have used the definition of $|\psi_V^{(j)}\rangle$ to obtain the equality.
By applying Eq.~(\ref{CPTPprop}) to Eq.~(\ref{statemp2.5}), the malicious prover's state at the end of Protocol 2 is
\begin{eqnarray}
\cfrac{1}{4}\sum_{\varphi^{(j)}}\mathcal{F}_1^{(j)}\left(|\psi_V^{(j)}\rangle\langle\psi_V^{(j)}|\right).
\end{eqnarray}
This quantum state can also be prepared at the end of step 1 in Protocol 1.

We next consider the case where a vacuum state is included in the input of the I1DC protocol.
Let the vacuum state be originated from the $k^\ast$th coherent state whose polarization angle is $\theta_{k^\ast}^{(j)}$.
Since the verifier selects each polarization angle independently, the malicious prover's state $\rho^{(j)}$ at the end of Protocol 2 does not depend on $\theta_{k^\ast}^{(j)}$ and hence $\varphi^{(j)}$.
Remember that $\varphi^{(j)}$ is just a variable replacement of $\theta_{k^\ast}^{(j)}$.
By using the CPTP map $\mathcal{F}_2^{(j)}$ that replaces any quantum state with the fixed state $\rho^{(j)}$, the prover's final state is
\begin{eqnarray}
\label{statemp3}
\mathcal{F}_2^{(j)}\left(|\psi_V^{(j)}\rangle\langle\psi_V^{(j)}|\right).
\end{eqnarray}
This quantum state can also be prepared at the end of step 1 in Protocol 1.

By combining the above arguments, when the deviation of the prover is independent in each repetition, the malicious prover's state after the $N$th repetition in Protocol 2 is
\begin{eqnarray}
&&\cfrac{1}{4^N}\sum_{\{\varphi^{(j)}\}_{j=1}^N}\bigotimes_{j=1}^N\mathcal{F}^{(j)}\left(|\psi_V^{(j)}\rangle\langle\psi_V^{(j)}|\right)\\
&=&\cfrac{1}{4^N}\sum_{\{\varphi^{(j)}\}_{j=1}^N}\left(\prod_{j=1}^N\mathcal{F}^{(j)}\right)\left(|\psi_V\rangle\langle\psi_V|\right),
\end{eqnarray}
where $\mathcal{F}^{(j)}\in\{\mathcal{F}_1^{(j)},\mathcal{F}_2^{(j)}\}$ for all $1\le j\le N$.
As shown in Appendix B, a similar argument holds even when the prover's attack is collective, i.e., the prover simultaneously handles all photons in all repetitions to deceive the verifier.
Therefore, $p_{{\rm acc}|{\rm (ii)}}\le 1-b/2$.
From this upper bound and Eqs.~(\ref{paccnew1}) and (\ref{paccmp}), when the prover is malicious,
\begin{eqnarray}
\label{paccmp2}
p_{\rm acc}\le1-\cfrac{b}{2}+N{\rm exp}\left(-\cfrac{me^{-2\alpha^2}\alpha^4}{2}\right).
\end{eqnarray}

In conclusion, from Eqs.~(\ref{pacchp}) and (\ref{paccmp2}) with $\alpha=1$ and $m=\lceil2e^2\log{(4N^2f(|x|))}\rceil$, where $\lceil\cdot\rceil$ is the ceiling function, the gap of $p_{\rm acc}$ between the honest and malicious provers' cases is
\begin{eqnarray}
\nonumber
&&1-\cfrac{a}{2}-N{\rm exp}\left(-\cfrac{me^{-2\alpha^2}\alpha^4}{2}\right)\\
\nonumber
&&-\left[1-\cfrac{b}{2}+N{\rm exp}\left(-\cfrac{me^{-2\alpha^2}\alpha^4}{2}\right)\right]\\
&\ge&\cfrac{b-a}{2}-\cfrac{1}{2Nf(|x|)}\\
\label{csgap}
&\ge&\cfrac{N-1}{2Nf(|x|)},
\end{eqnarray}
which is the inverse of a polynomial in $|x|$.
Thus, our physically classical verification protocol efficiently distinguishes the honest and malicious provers.

\section{Conclusion \& discussion}
We have proposed an efficient verification protocol that removes the necessity of qubit communication.
Since all apparatuses required for the verifier are a telecom-band laser with linear optical elements and a classical computer, our results would facilitate the realization of verifiable quantum computers.

To further improve the practicality of our protocol, we discuss the phase randomization implemented for the coherent state.
Several papers (e.g., Ref.~\cite{CZLM15}) have pointed out the difficulty of the continuous phase randomization.
By calculating the fidelity between the continuous-phase-randomized (see Eq.~(\ref{cprcs})) and discrete-phase-randomized coherent states, we evaluate how many classical bits would be required for the phase randomization.
For simplicity, let
\begin{eqnarray}
\rho_{\theta_k^{(j)}}^\infty\equiv e^{-\alpha^2}\sum_{n=0}^\infty\cfrac{\alpha^{2n}}{n!}|n_{\theta_k^{(j)}}\rangle\langle n_{\theta_k^{(j)}}|
\end{eqnarray}
and
\begin{eqnarray}
\rho_{\theta_k^{(j)}}^R\equiv\cfrac{1}{R}\sum_{j=0}^{R-1}|{e^{i2j\pi/R}\alpha}_{\theta_k^{(j)}}\rangle\langle{e^{i2j\pi/R}\alpha}_{\theta_k^{(j)}}|
\end{eqnarray}
be the continuous-phase-randomized and discrete-phase-randomized coherent states, where $|\beta_\theta\rangle\equiv e^{-|\beta|^2/2}\sum_{n=0}^\infty(\beta^n/\sqrt{n!})|n_\theta\rangle$ is the phase-fixed coherent state with the polarization angle $\theta$ for any complex number $\beta$.
We denote the fidelity between two quantum states $\rho$ and $\sigma$ as $F(\rho,\sigma)$.
From Ref.~\cite{JYWCGH24}, when $\alpha=1$ and $R\ge e^2+1$, the fidelity is
\begin{eqnarray}
\nonumber
&&F\left(\bigotimes_{j,k}\left(\rho_{\theta_k^{(j)}}^\infty\otimes|\theta_k^{(j)}\rangle\langle\theta_k^{(j)}|\right),\bigotimes_{j,k}\left(\rho_{\theta_k^{(j)}}^R\otimes|\theta_k^{(j)}\rangle\langle\theta_k^{(j)}|\right)\right)\\
&=&F\left(\rho_{\theta_k^{(j)}}^\infty,\rho_{\theta_k^{(j)}}^R\right)^{mN}\\
&=&\left\{\sum_{j=0}^{R-1}\sqrt{\sum_{k=0}^\infty\left[\cfrac{e^{-1}}{(kR+j)!}\right]^2}\right\}^{2mN}\\
&\ge&e^{-2mN}\left(\sum_{j=0}^{R-1}\cfrac{1}{j!}\right)^{2mN}\\
&\ge&e^{-2mN}\left[e-\cfrac{e}{(R-1)!}\right]^{2mN}\\
&\ge&1-2mN\left(\cfrac{e}{R-1}\right)^{R-1}\equiv F_{\rm min}.
\end{eqnarray}
Let $p_{\rm acc}$ and $q_{\rm acc}$ be the acceptance probabilities of our physically classical verification protocol with continuous-phase-randomized and discrete-phase-randomized coherent states, respectively.
Eq.~(\ref{csgap}) implies that $|p_{\rm acc}-q_{\rm acc}|\le(N-1)/[4Nf(|x|)]$ is sufficient for our protocol to correctly distinguish between honest and malicious provers.
From Ref.~\cite{HQ12} that studies the fidelity and trace distance for any quantum states in an infinite-dimensional separable complex Hilbert space, $|p_{\rm acc}-q_{\rm acc}|\le\sqrt{1-F_{\rm min}}$, and hence using $R\ge e^2+1$ results in $R=\lceil\log{[32mN^3f(|x|)^2/(N-1)^2]}+1\rceil$.
Since $m=\lceil2e^2\log{(4N^2f(|x|))}\rceil$, this calculation shows that a log of log number of classical bits is sufficient for the phase randomization.

The RBSP protocol~\cite{DKL12} was improved or modified in Refs.~\cite{ZL16,ZL18,NHS19,PYZMDZLW22}.
As another direction to improve our protocol, it would be interesting to consider their applicability to the verification of quantum computation.
Furthermore, as with the implementation security~\cite{ZNC24} of QKD, it would be important to devise the verification protocols under several imperfections such as channel loss and noises and the correlation between coherent states.
An efficient way~\cite{KLMO24} that removes trusted quantum state preparations and measurements from verification protocols may be useful for this purpose.

In this paper, we have proposed a physically classical verification protocol by combining the protocols in Refs.~\cite{M20} and \cite{DKL12}.
On the other hand, there are compilers that make blind quantum computing protocols verifiable~\cite{M18,KKLMO22}.
Blind quantum computation is a secure protocol such that a user can delegate universal quantum computation to a remote quantum computer without disclosing the user's input, quantum algorithm, and output.
Although Ref.~\cite{M18} seems to implicitly assume perfect blindness (i.e., perfect security), the blind quantum computing protocol with coherent states proposed in Ref.~\cite{DKL12} does not have perfect blindness.
This is why it could not be directly applied to Ref.~\cite{DKL12} to obtain a physically classical verification protocol.
When we apply Ref.~\cite{KKLMO22} to Ref.~\cite{DKL12}, the resultant protocol should require a polynomial number of communication rounds between the verifier and prover, while our protocol is one round because $N$ repetitions in Protocol 2 can be done in parallel, and the prover can send $\{n_k^{(j)}\}_{k,j}$ and $\{w,z\}_{w,z}$ simultaneously.
Our protocol successfully reveals an advantage of the transmission of coherent light in the sense that if the ultimate protocol in Fig.~\ref{summary} with a constant number of rounds can be constructed, then ${\sf BQP}$ is contained in the third level of the polynomial hierarchy~\cite{FHM18}.
This set containment is considered to be unlikely due to an oracle separation between ${\sf BQP}$ and ${\sf PH}$~\cite{RT22}.

It would be interesting to devise physically classical verification protocols by modifying other existing verification protocols (e.g., Ref.~\cite{B18}) and then combining them with the RBSP protocol~\cite{DKL12}.

\section*{ACKNOWLEDGMENTS}
We thank Seiichiro Tani and Tomoyuki Morimae for helpful discussions.
YT is partially supported by JST [Moonshot R\&D -- MILLENNIA Program] Grant Number JPMJMS2061.
AM is partially supported by JST, ACT-X Grant No. JPMJAX210O, Japan.

\setcounter{section}{0}
\clearpage
\widetext
\section*{Appendix A: Acceptance probabilities in Protocol 1}
The calculation is based on the idea in Ref.~\cite{MT20}.
We first consider the case where the prover is honest.
Let $h\equiv h_1\ldots h_N$, $s\equiv s_1\ldots s_N$, $\bar{\alpha}\equiv\alpha\oplus 1$ for any classical bit $\alpha\in\{0,1\}$, and $\delta_{\alpha\beta}$ be the Kronecker delta such that it is equal to one or zero when $\alpha=\beta$ or $\alpha\neq\beta$ for the two classical bits $\alpha\in\{0,1\}$ and $\beta\in\{0,1\}$, respectively.
Since there exists the low-energy state $|\eta\rangle$, and $\{\Pi_{wz}\}_{w,z}$ corresponds to the $N$ parallel measurements in the Bell basis $\{|\phi_{\alpha\beta}\rangle\equiv(Z^\beta X^\alpha\otimes I)(|00\rangle+|11\rangle)/\sqrt{2}\}_{\alpha,\beta\in\{0,1\}}$, the acceptance probability $p_{\rm acc}$ is
\begin{eqnarray}
\nonumber
&&\cfrac{1}{2^{2N}}\sum_{h,s\in\{0,1\}^N}\sum_{w,z\in\{0,1\}^N}\left(\bigotimes_{k=1}^N\langle\phi_{w_kz_k}|\right)\left(|\eta\rangle\langle\eta|\otimes|\psi_V\rangle\langle\psi_V|\right)\left(\bigotimes_{k=1}^N|\phi_{w_kz_k}\rangle\right)\left\{\sum_{i<j}p_{ij}^{(x)}\left[\delta_{\bar{h_i}h_j}+\delta_{h_ih_j}\cfrac{1-c_{ij}^{(x)}(-1)^{s'_i+s'_j}}{2}\right]\right\}\\
\\
&=&\cfrac{1}{2}+\cfrac{1}{2^{2N}}\sum_{h,s,w,z\in\{0,1\}^N}\left(\bigotimes_{k=1}^N\langle\phi_{w_kz_k}|\right)\left(|\eta\rangle\langle\eta|\otimes|\psi_V\rangle\langle\psi_V|\right)\left(\bigotimes_{k=1}^N|\phi_{w_kz_k}\rangle\right)\left[\sum_{i<j}p_{ij}^{(x)}\delta_{h_ih_j}\cfrac{1-c_{ij}^{(x)}(-1)^{s'_i+s'_j}}{2}\right]\\
&=&\cfrac{1}{2}+\cfrac{1}{2^{3N}}\sum_{h,s,w,z\in\{0,1\}^N}\left(\bigotimes_{k=1}^N\langle s_k|H_kS_k^{h_k}X_k^{w_k}Z_k^{z_k}\right)|\eta\rangle\langle\eta|\left(\bigotimes_{k=1}^NZ_k^{z_k}X_k^{w_k}{S_k^\dag}^{h_k}H_k|s_k\rangle\right)\left[\sum_{i<j}p_{ij}^{(x)}\delta_{h_ih_j}\cfrac{1-c_{ij}^{(x)}(-1)^{s'_i+s'_j}}{2}\right]\\
\nonumber
&=&\cfrac{1}{2}+\cfrac{1}{2^{3N}}\sum_{h,s,w,z\in\{0,1\}^N}\left(\bigotimes_{k=1}^N\langle s_k|H_kS_k^{h_k}X_k^{w_k}Z_k^{z_k}\right)|\eta\rangle\langle\eta|\left(\bigotimes_{k=1}^NZ_k^{z_k}X_k^{w_k}{S_k^\dag}^{h_k}\right)\times\\
&&\left[\sum_{i<j}p_{ij}^{(x)}\delta_{h_ih_j}\left(\bigotimes_{k=1}^NZ_k^{z_k+h_kw_k}\right)\cfrac{I^{\otimes N}-c_{ij}^{(x)}X_i\otimes X_j}{2}\left(\bigotimes_{k=1}^NZ_k^{z_k+h_kw_k}\right)\right]\left(\bigotimes_{k=1}^NH_k|s_k\rangle\right)\\
\nonumber
&=&\cfrac{1}{2}+\cfrac{1}{2^{3N}}\sum_{h,w,z\in\{0,1\}^N}\sum_{i<j}p_{ij}^{(x)}\delta_{h_ih_j}\times\\
&&{\rm Tr}\left[\left(\bigotimes_{k=1}^NS_k^{h_k}X_k^{w_k}Z_k^{z_k}\right)|\eta\rangle\langle\eta|\left(\bigotimes_{k=1}^NZ_k^{z_k}X_k^{w_k}{S_k^\dag}^{h_k}\right)\left(\bigotimes_{k=1}^NZ_k^{z_k+h_kw_k}\right)\cfrac{I^{\otimes N}-c_{ij}^{(x)}X_i\otimes X_j}{2}\left(\bigotimes_{k=1}^NZ_k^{z_k+h_kw_k}\right)\right]\\
\nonumber
&=&\cfrac{1}{2}+\cfrac{1}{16}\sum_{i<j}\sum_{h_i,w_i,w_j\in\{0,1\}}p_{ij}^{(x)}\times\\
&&{\rm Tr}\left[|\eta\rangle\langle\eta|\left(X_i^{w_i}{S_i^\dag}^{h_i}\otimes X_j^{w_j}{S_j^\dag}^{h_i}\right)\left(Z_i^{h_iw_i}\otimes Z_j^{h_iw_j}\right)\cfrac{I^{\otimes N}-c_{ij}^{(x)}X_i\otimes X_j}{2}\left(Z_i^{h_iw_i}\otimes Z_j^{h_iw_j}\right)\left(S_i^{h_i}X_i^{w_i}\otimes S_j^{h_i}X_j^{w_j}\right)\right]\\
&=&\cfrac{1}{2}+\cfrac{1}{2}{\rm Tr}\left[|\eta\rangle\langle\eta|\sum_{i<j}\cfrac{p_{ij}^{(x)}}{2}\left(\cfrac{I^{\otimes N}-c_{ij}^{(x)}X_i\otimes X_j}{2}+\cfrac{I^{\otimes N}-c_{ij}^{(x)}Y_i\otimes Y_j}{2}\right)\right]\\
&=&\cfrac{1+\langle\eta|\left(I^{\otimes N}-H_x\right)|\eta\rangle}{2}\ge 1-\cfrac{a}{2}.
\end{eqnarray}

We next consider the case where the prover is malicious.
The acceptance probability $p_{\rm acc}$ is
\begin{eqnarray}
&&\cfrac{1}{2^{2N}}\sum_{h,s\in\{0,1\}^N}\sum_{w,z\in\{0,1\}^N}\langle\psi_V|\Pi_{wz}|\psi_V\rangle\left\{\sum_{i<j}p_{ij}^{(x)}\left[\delta_{\bar{h_i}h_j}+\delta_{h_ih_j}\cfrac{1-c_{ij}^{(x)}(-1)^{s'_i+s'_j}}{2}\right]\right\}\\
&=&\cfrac{1}{2}+\cfrac{1}{2^{2N}}\sum_{h,s,w,z\in\{0,1\}^N}\langle\psi_V|\Pi_{wz}|\psi_V\rangle\left\{\sum_{i<j}p_{ij}^{(x)}\delta_{h_ih_j}\cfrac{1-c_{ij}^{(x)}(-1)^{s'_i+s'_j}}{2}\right\}\\
\nonumber
&=&\cfrac{1}{2}+\cfrac{1}{2^{2N}}\sum_{h,s,w,z\in\{0,1\}^N}\left(\bigotimes_{k=1}^N\langle s_k|H_k{S_k^\dag}^{h_k}\right)\Pi_{wz}\left(\bigotimes_{k=1}^N S_k^{h_k}\right)\times\\
&&\left[\sum_{i<j}p_{ij}^{(x)}\delta_{h_ih_j}\left(\bigotimes_{k=1}^NZ_k^{z_k+h_kw_k}\right)\cfrac{I^{\otimes N}-c_{ij}^{(x)}X_i\otimes X_j}{2}\left(\bigotimes_{k=1}^NZ_k^{z_k+h_kw_k}\right)\right]\left(\bigotimes_{k=1}^NH_k|s_k\rangle\right)\\
\nonumber
&=&\cfrac{1}{2}+\cfrac{1}{2^{2N}}\sum_{h,w,z\in\{0,1\}^N}\times\\
&&{\rm Tr}\left[\Pi_{wz}\left[\sum_{i<j}p_{ij}^{(x)}\delta_{h_ih_j}\left(\bigotimes_{k=1}^NZ_k^{z_k+h_kw_k}\right)\left(\bigotimes_{k=1}^N S_k^{h_k}\right)\cfrac{I^{\otimes N}-c_{ij}^{(x)}X_i\otimes X_j}{2}\left(\bigotimes_{k=1}^N{S_k^\dag}^{h_k}\right)\left(\bigotimes_{k=1}^NZ_k^{z_k+h_kw_k}\right)\right]\right]\\
\nonumber
&=&\cfrac{1}{2}+\cfrac{1}{2^{N+2}}\sum_{i<j}p_{ij}^{(x)}\sum_{h_i\in\{0,1\}}\sum_{w,z\in\{0,1\}^N}{\rm Tr}\left[\left(\bigotimes_{k=1}^NX_k^{w_k}Z_k^{z_k}\right)\Pi_{wz}\left(\bigotimes_{k=1}^NZ_k^{z_k}X_k^{w_k}\right)\times\right.\\
&&\left.\left[\left(X_i^{w_i}Z_i^{h_iw_i}\otimes X_j^{w_j}Z_j^{h_iw_j}\right)\left(S_i^{h_i}\otimes S_j^{h_i}\right)\cfrac{I^{\otimes N}-c_{ij}^{(x)}X_i\otimes X_j}{2}\left({S_i^\dag}^{h_i}\otimes {S_j^\dag}^{h_i}\right)\left(Z_i^{h_iw_i}X_i^{w_i}\otimes Z_j^{h_iw_j}X_j^{w_j}\right)\right]\right]\\
&=&\cfrac{1}{2}+\cfrac{1}{2}{\rm Tr}\left[\left[\cfrac{1}{2^N}\sum_{w,z\in\{0,1\}^N}\left(\bigotimes_{k=1}^NX_k^{w_k}Z_k^{z_k}\right)\Pi_{wz}\left(\bigotimes_{k=1}^NZ_k^{z_k}X_k^{w_k}\right)\right]\left(I^{\otimes N}-H_x\right)\right]\le 1-\cfrac{b}{2},
\end{eqnarray}
where we have used the observation that $[\sum_{w,z\in\{0,1\}^N}(\otimes_{k=1}^NX_k^{w_k}Z_k^{z_k})\Pi_{wz}(\otimes_{k=1}^NZ_k^{z_k}X_k^{w_k})]/2^N$ is a quantum state to obtain the last inequality.

\section*{Appendix B: Malicious prover's final state in Protocol 2 for collective attacks}
In this appendix, we derive the malicious prover's quantum state after the $N$th repetition in Protocol 2 under any collective attack.
Since we consider case (ii), we can assume that the number $\tilde{m}_0^{(j)}$ of vacuum states is less than $m_0^{(j)}+m_1^{(j)}$ for all $1\le j\le N$.
Therefore, the I1DC protocol in the $j$th repetition has a state whose actual photon number is zero or one as an input.
Let it be the $l^\ast_j$th input state whose polarization angle is $\sigma_{l^\ast_j}^{(j)}$.
For simplicity, we define $\vec{\sigma}\in\{0,\pi/2,\pi,3\pi/2\}^{(m-1)N}$ as the string of all the polarization angles except for $\{\sigma_{l^\ast_j}^{(j)}\}_{j=1}^N$.
In general, the prover's final state can be written as
\begin{eqnarray}
\cfrac{1}{4^{mN}}\sum_{\sigma_{l^\ast_1}^{(1)},\sigma_{l^\ast_2}^{(2)},\ldots,\sigma_{l^\ast_N}^{(N)},\vec{\sigma}}\sum_{\vec{o}^{\ (1)},\vec{o}^{\ (2)},\ldots,\vec{o}^{\ (N)}}p_{\sigma_{l^\ast_1}^{(1)},\sigma_{l^\ast_2}^{(2)},\ldots,\sigma_{l^\ast_N}^{(N)},\vec{\sigma}}(\vec{o}^{\ (1)},\vec{o}^{\ (2)},\ldots,\vec{o}^{\ (N)})\mathcal{E}_{\vec{o}^{\ (1)},\vec{o}^{\ (2)},\ldots,\vec{o}^{\ (N)},\vec{\sigma}}\left(\bigotimes_{j=1}^N|+_{\sigma_{l^\ast_j}^{(j)}}\rangle\langle +_{\sigma_{l^\ast_j}^{(j)}}|\right),\ \ \
\end{eqnarray}
where $p_{\sigma_{l^\ast_1}^{(1)},\sigma_{l^\ast_2}^{(2)},\ldots,\sigma_{l^\ast_N}^{(N)},\vec{\sigma}}(\vec{o}^{\ (1)},\vec{o}^{\ (2)},\ldots,\vec{o}^{\ (N)})$ is the probability of outputting $\{\vec{o}^{\ (j)}\}_{j=1}^N$, and $\mathcal{E}_{\vec{o}^{\ (1)},\vec{o}^{\ (2)},\ldots,\vec{o}^{\ (N)},\vec{\sigma}}$ is a quantum operation to be applied when the measurement outcomes are $\{\vec{o}^{\ (j)}\}_{j=1}^N$.
The subscript $\sigma_{l^\ast_1}^{(1)},\sigma_{l^\ast_2}^{(2)},\ldots,\sigma_{l^\ast_N}^{(N)}$ of $p_{\sigma_{l^\ast_1}^{(1)},\sigma_{l^\ast_2}^{(2)},\ldots,\sigma_{l^\ast_N}^{(N)},\vec{\sigma}}(\vec{o}^{\ (1)},\vec{o}^{\ (2)},\ldots,\vec{o}^{\ (N)})$ just represents that the probability can depend on each $|+_{\sigma_{l^\ast_j}^{(j)}}\rangle$ if the $l^\ast_j$th input qubit of the I1DC protocol in the $j$th repetition is not a vacuum state; it does not mean that the prover's CPTP map is constructed by using the values of $\{\sigma_{l^\ast_j}^{(j)}\}_{j=1}^N$.
The other subscript, $\vec{\sigma}$, of $p_{\sigma_{l^\ast_1}^{(1)},\sigma_{l^\ast_2}^{(2)},\ldots,\sigma_{l^\ast_N}^{(N)},\vec{\sigma}}(\vec{o}^{\ (1)},\vec{o}^{\ (2)},\ldots,\vec{o}^{\ (N)})$ and $\mathcal{E}_{\vec{o}^{\ (1)},\vec{o}^{\ (2)},\ldots,\vec{o}^{\ (N)},\vec{\sigma}}$ means that they may depend on $\vec{\sigma}$ because the corresponding coherent states may include more than one photon.
Since the verifier keeps the values of $\{\sigma_{l^\ast_j}^{(j)}\}_{j=1}^N$ private and the actual photon number of each $|+_{\sigma_{l^\ast_j}^{(j)}}\rangle$ is one or zero, as with Eq.~(\ref{CPTPprop}), there exists a CPTP map $\mathcal{F}$ such that
\begin{eqnarray}
\nonumber
&&\mathcal{F}\left(\bigotimes_{j=1}^N|+_{\sigma_{l^\ast_j}^{(j)}}\rangle\langle +_{\sigma_{l^\ast_j}^{(j)}}|\right)\\
\label{CPTPca}
&=&\cfrac{1}{4^{(m-1)N}}\sum_{\vec{\sigma}}\sum_{\vec{o}^{\ (1)},\vec{o}^{\ (2)},\ldots,\vec{o}^{\ (N)}}p_{\sigma_{l^\ast_1}^{(1)},\sigma_{l^\ast_2}^{(2)},\ldots,\sigma_{l^\ast_N}^{(N)},\vec{\sigma}}(\vec{o}^{\ (1)},\vec{o}^{\ (2)},\ldots,\vec{o}^{\ (N)})\mathcal{E}_{\vec{o}^{\ (1)},\vec{o}^{\ (2)},\ldots,\vec{o}^{\ (N)},\vec{\sigma}}\left(\bigotimes_{j=1}^N|+_{\sigma_{l^\ast_j}^{(j)}}\rangle\langle +_{\sigma_{l^\ast_j}^{(j)}}|\right)
\end{eqnarray}
for any $\{\sigma_{l^\ast_j}^{(j)}\}_{j=1}^N$.
Recall that when the actual photon number of $|+_{\sigma_{l^\ast_j}^{(j)}}\rangle$ is zero, the probability $p_{\sigma_{l^\ast_1}^{(1)},\sigma_{l^\ast_2}^{(2)},\ldots,\sigma_{l^\ast_N}^{(N)},\vec{\sigma}}(\vec{o}^{\ (1)},\vec{o}^{\ (2)},\ldots,\vec{o}^{\ (N)})$ does not depend on $\sigma_{l^\ast_j}^{(j)}$.

For any $\vec{\sigma}$ and $\{\vec{o}^{\ (j)}\}_{j=1}^N$, there exist $\{\theta_j\}_{j=1}^N\in\{0,\pi/2,\pi,3\pi/2\}^N$ and $\{c_j\}_{j=1}^N\in\{1,-1\}^N$ such that
\begin{eqnarray}
\label{vr2}
\varphi^{(j)}=\theta_j+c_j\sigma_{l^\ast_j}^{(j)}
\end{eqnarray}
for all $j$.
It is worth mentioning that when the value of $\sigma_{l^\ast_j}^{(j)}$ is chosen from $\{0,\pi/2,\pi,3\pi/2\}$ uniformly at random, the value of $\varphi^{(j)}$ is also  determined uniformly at random on the same range.
Therefore, from Eqs.~(\ref{vr2}) and (\ref{CPTPca}), the prover's final state is
\begin{eqnarray}
\nonumber
&&\cfrac{1}{4^{mN}}\sum_{\sigma_{l^\ast_1}^{(1)},\sigma_{l^\ast_2}^{(2)},\ldots,\sigma_{l^\ast_N}^{(N)},\vec{\sigma}}\sum_{\vec{o}^{\ (1)},\vec{o}^{\ (2)},\ldots,\vec{o}^{\ (N)}}p_{\sigma_{l^\ast_1}^{(1)},\sigma_{l^\ast_2}^{(2)},\ldots,\sigma_{l^\ast_N}^{(N)},\vec{\sigma}}(\vec{o}^{\ (1)},\vec{o}^{\ (2)},\ldots,\vec{o}^{\ (N)})\mathcal{E}_{\vec{o}^{\ (1)},\vec{o}^{\ (2)},\ldots,\vec{o}^{\ (N)},\vec{\sigma}}\left(\bigotimes_{j=1}^N|+_{\sigma_{l^\ast_j}^{(j)}}\rangle\langle +_{\sigma_{l^\ast_j}^{(j)}}|\right)\\
\nonumber
&=&\cfrac{1}{4^{mN}}\sum_{\varphi^{(1)},\varphi^{(2)},\ldots,\varphi^{(N)},\vec{\sigma}}\sum_{\vec{o}^{\ (1)},\vec{o}^{\ (2)},\ldots,\vec{o}^{\ (N)}}p_{\varphi^{(1)},\varphi^{(2)},\ldots,\varphi^{(N)},\vec{\sigma}}(\vec{o}^{\ (1)},\vec{o}^{\ (2)},\ldots,\vec{o}^{\ (N)})\mathcal{E}_{\vec{o}^{\ (1)},\vec{o}^{\ (2)},\ldots,\vec{o}^{\ (N)},\vec{\sigma}}\left(\bigotimes_{j=1}^N|+_{\varphi^{(j)}}\rangle\langle +_{\varphi^{(j)}}|\right)\\
\\
&=&\cfrac{1}{4^N}\sum_{\varphi^{(1)},\varphi^{(2)},\ldots,\varphi^{(N)}}\mathcal{F}\left(\bigotimes_{j=1}^N|+_{\varphi^{(j)}}\rangle\langle +_{\varphi^{(j)}}|\right)\\
&=&\cfrac{1}{4^N}\sum_{\varphi^{(1)},\varphi^{(2)},\ldots,\varphi^{(N)}}\mathcal{F}\left(|\psi_V\rangle\langle\psi_V|\right).
\end{eqnarray}
Since this quantum state can be prepared at the end of step 1 in Protocol 1, the inequality $p_{{\rm acc}|{\rm (ii)}}\le 1-b/2$ holds.

\begin{thebibliography}{99}
\bibitem{FN1}There is another approach based on the in-class interactive proof~\cite{AG17,GKL21}, but we do not focus on it in this paper because it is currently not practical in the following sense.
Although this type of verification protocols is constructed by weakening the prover's computational ability in the interactive proof system for ${\sf BQP}$, superpolynomial-time quantum computation is still necessary for the prover to be accepted by the verifier.
It is worth mentioning that this approach is promising because the ultimate goal is achieved if the required prover's ability can be further weakened.
Furthermore, in this paper, we do not focus on the verification of subuniversal quantum computing models, such as instantaneous quantum polynomial time (IQP)~\cite{BMS16,KD19,BCJ23}, deterministic quantum computation with $1$ pure qubit (DQC1)~\cite{FKMNTT18,KKD14}, boson sampling~\cite{AA13,TMBM14,AA14,AVFSCOWS19,CGKM21}, and noisy intermediate-scale quantum (NISQ) computation~\cite{P18,TTMT22}.
\bibitem{FN2}In Ref.~\cite{M20}, the second term of the Hamiltonian $H_x$ is $(I^{\otimes N}+c_{ij}^{(x)}Z_i\otimes Z_j)/2$.
The transformation from the original Hamiltonian $H'_x$ to our Hamiltonian $H_x$ is straightforward.
Let $V\equiv YSHS^\dag$ for the Pauli-$Y$ operator $Y$, the $S$ gate $S\equiv|0\rangle\langle 0|+i|1\rangle\langle 1|$, and the Hadamard gate $H\equiv|+\rangle\langle 0|+|-\rangle\langle 1|$.
It is trivial to show $VZV^\dag=Y$ and $VXV^\dag=X$, and hence our Hamiltonian $H_x$ can be obtained by applying $V^{\otimes N}$ to the original Hamiltonian $H'_x$.
When the prover is honest, our low-energy state $|\eta\rangle$ can be obtained by applying $V^{\otimes N}$ to the efficiently preparable low-energy state of the original Hamiltonian $H'_x$. On the other hand, when the prover is malicious, if there exists a quantum state $\rho$ whose energy is lower than $b$ for $H_x$, then ${\rm Tr}[H'_x{V^\dag}^{\otimes N}\rho V^{\otimes N}]<b$, which contradicts the fact that the ground-state energy of $H'_x$ is at least $b$.
\bibitem{FNM1}For example, when $n_1^{(j)}=2$, $n_2^{(j)}=n_3^{(j)}=0$, $n_4^{(j)}=1$, and $n_5^{(j)}=n_6^{(j)}=n_7^{(j)}=n_8^{(j)}=0$ with $m=8$, the honest prover keeps the state $|1_{\theta_1^{(j)}}\rangle|1_{\theta_4^{(j)}}\rangle$ at hand.
This is consistent with the fact that $m-m_0^{(j)}=8-6=2$.
\bibitem{FNM3}Since vacuum states are neglected in step (b) of Protocol 2, the $l$th remaining photon may not be extracted from the $l$th coherent state in step (a) of Protocol 2.
In other words, the polarization angle of the $l$th remaining photon is, in general, not $\theta_l^{(j)}$.
This is why we introduce a different notation $\sigma_l^{(j)}$.
\bibitem{FNM2}Just in case, we mention that to use the Hoeffding inequality, we define a random variable that takes the value one and zero when the photon number is zero or one and is more than one, respectively.
\end{thebibliography}
\end{document}